\documentclass[twocolumn,showpacs,preprintnumbers,amsmath,amssymb]{revtex4}

\usepackage{graphicx}
\usepackage{dcolumn}
\usepackage{bm}
\draft

\begin{document}
\def\bsv{{\hbox{\boldmath $v$}}}
\def\bsk{{\hbox{\boldmath $k$}}}

\title{Pion Interferometry for a Granular Source of Quark-Gluon Plasma 
Droplets}

\author{W. N. Zhang$^{1,2}$}
\author{M. J. Efaaf$^1$}
\author{Cheuk-Yin Wong$^{2,3}$}

\affiliation{
$^1$Department of Physics, Harbin Institute of Technology, 
Harbin, 150006, P. R. China\\
$^2$Physics Division, Oak Ridge National Laboratory, Oak Ridge, TN
37831, U.S.A.\\
$^3$Department of Physics, University of Tennessee, Knoxville, TN
37996, U.S.A.
}

\date{\today}

\begin{abstract}
We examine the two-pion interferometry for a granular source of
quark-gluon plasma droplets. The evolution of the droplets is
described by relativistic hydrodynamics with an equation of state
suggested by lattice gauge results.  Pions are assumed to
be emitted thermally from the droplets at the freeze-out configuration
characterized by a freeze-out temperature $T_f$.  We find that the HBT
radius $R_{\rm out}$ decreases if the initial size of the droplets
decreases.  On the other hand, $R_{\rm side}$ depends on the droplet
spatial distribution and is relatively independent of the droplet
size.  It increases with an increase in the width of the spatial
distribution and the collective-expansion velocity of the droplets.
As a result, the value of $R_{\rm out}$ can lie close to $R_{\rm
side}$ for a granular quark-gluon plasma source.  The granular model
of the emitting source may provide an explanation to the RHIC HBT
puzzle and may lead to a new insight into the dynamics of the
quark-gluon plasma phase transition.
\end{abstract}


\maketitle

Recent experimental pion HBT measurements at RHIC give the ratio of
$R_{\rm out}/ R_{\rm side} \approx 1$ \cite{PHE02,STA01} which is
contrary to many theoretical expectations
\cite{PHE02,Ris96,Tea99,Wie99,Sof01,Wei02,Pra03}.  This
RHIC HBT puzzle hints that the pion emitting time may be very short
\cite{Sof02,Mol02}.  Various models have been put forth to explain the
HBT puzzle \cite{Sof02a,Hei02,Lin02,Tea03,Cso03,Mol04}.

In this paper, we propose a simple model of a granular source of
quark-gluon plasma droplets \cite{Sei89,Kaj91,Pra92,Zha95,Zha00} to
explain the puzzle.  The possible occurrence of a granular structure
of droplets during a first-order QCD phase transition was discussed by
Witten \cite{Wit84} and examined by many authors
\cite{Cse92,Ven94,Ala99,Cse03,Ran04}.  We assume that pions are
emitted thermally from hydrodynamically expanding droplets at a
freeze-out temperature $T_f$ and use relativistic hydrodynamics to
describe the evolution of quark-gluon plasma droplets with an equation
of state suggested by QCD lattice gauge results
\cite{Ris96,Ris98,Lae96,Bla87}.  The two-pion correlation function can
then be calculated after knowing the hydrodynamical solution
\cite{Zha95,Zha00,Zha04}.  As the average freeze-out time is
approximately proportional to the initial droplet size, we would like
to see whether a granular source with many smaller droplets and their
corresponding smaller freeze-out times will lead to $R_{\rm out}$
close to $R_{\rm side}$.

We study a quark-gluon plasma with no net-baryon content and use a 
equation of state of the fluid in terms of the entropy density 
function $s(T)$ by 
\cite{Ris96,Ris98,Lae96,Bla87}
\begin{equation}
\label{eos}
{s (T) \over {s_c}} = \Bigg[{T\over {T_c}}\Bigg]^3 \Bigg(1+{{d_Q-d_H}
\over {d_Q+d_H}}\tanh\Bigg[{{T-T_c}\over {\Delta T}}\Bigg] \Bigg) \,.
\end{equation}
Here $d_Q$ and $d_H$ are respectively the degrees of freedom in the
quark-gluon plasma phase and the hadronic phase, $T_c \approx 160$ MeV
is the transition temperature, $s_c$ is the entropy density at $T_c$, 
and $\Delta T$ (between 0 and $0.1T_c$) is the width of the transition
\cite{Ris96,Ris98}.  In this paper, we take $d_Q=37$, $d_H=3$, $T_c=160$
MeV as in Ref. \cite{Ris96,Ris98}, and take $\Delta T = 0.05T_c$.

We shall make the approximation that the hydrodynamical solution for
many independent droplets can be obtained by superposing the
hydrodynamical solution of a single droplet. It suffices to focus
attention on the hydrodynamics of a single droplet. Knowing the
entropy density $s(T)$, one can get the pressure $p$, energy density
$\epsilon$, and the velocity of sound $c_{\rm S}$ in the droplet with
the following equations as in Ref. \cite{Ris96,Ris98},
\begin{equation}
\label{pecs}
p=\int_0^T dT' s(T'), ~~~~~~\epsilon = Ts-p, ~~~~~
c_{\rm S}^2 = {{dp}\over{d\epsilon}} \,.
\end{equation}

The energy momentum tensor of a thermalized fluid cell in the 
center-of-mass frame of the droplet is \cite{Ris96,Lan59,Kol03}
\begin{equation}
\label{tensor}
T^{\mu \nu} (x) = \big [ \epsilon(x) + p(x) \big ] u^{\mu}(x) u^{\nu}(x)
- p(x) g^{\mu \nu} \, ,
\end{equation}
where $x$ is the space-time coordinate, $u^{\mu}=\gamma (1,\bsv)$
is the 4-velocity of the cell, and $g^{\mu \nu}$ is the metric tensor.
With the local conservation of energy and momentum, one can follow
Rischke and Gyulassy and get the equations for spherical geometry as
\cite{Ris96,Ris98}
\begin{eqnarray} 
\label{eqe}
\partial_t E + \partial_r [(E+p)v] & = &
- F \, , \\
\label{eqm}
\partial_t M + \partial_r (Mv+p) & = &
- G \, ,
\end{eqnarray}
where $E \equiv T^{00}$, $M \equiv T^{0r}$, 
\begin{eqnarray} 
\label{FGR}
F = \frac{2 v}{r} (E+p),~~~~~~~G = \frac{2 v}{r} M .
\end{eqnarray} 
We assume the initial conditions as \cite{Ris96,Ris98}
\begin{eqnarray}
\label{inic}
\epsilon(0,r) = \left\{
           \begin{array}{ll}
           \epsilon_0, & r<r_d, \\
           0, & r>r_d,
           \end{array}  \right. ~~~
v(0,r) = \left\{
           \begin{array}{ll}
           0, & r<r_d, \\
           1, & r>r_d,
           \end{array}  \right. 
\end{eqnarray}
where $\epsilon_0 = 1.875 T_c s_c$ \cite{Ris96,Ris98} is the initial
energy density of the droplets, and $r_d$ is the initial droplet
radius.  Using the Harten-Lax-van Leer-Einfeldt (HLLE) scheme
\cite{Ris96,Ris98,Sch93,Har83} and the relation of $p=p(\epsilon)$
obtained from Eqs. (\ref{eos}) and (\ref{pecs}), one can get the
solution of the hydrodynamical equations for $F=G=0$.  One then
obtains the solution for Eqs. (\ref{eqe}) and (\ref{eqm}) by using the
Sod's operator splitting method \cite{Ris96,Ris98,Sod77}.  The grid
spacing for the HLLE scheme is taken to be $\Delta x=0.01r_d$, and the
time step for the HLLE scheme and Sod's method corrector is $\Delta t
= 0.99\Delta x$ \cite{Ris96,Ris98}.  Figure 1(a) and 1(b) show the
temperature and velocity profiles of the droplet.  Figure 1(c) gives
the isotherms for the droplet.

\begin{figure}
\includegraphics{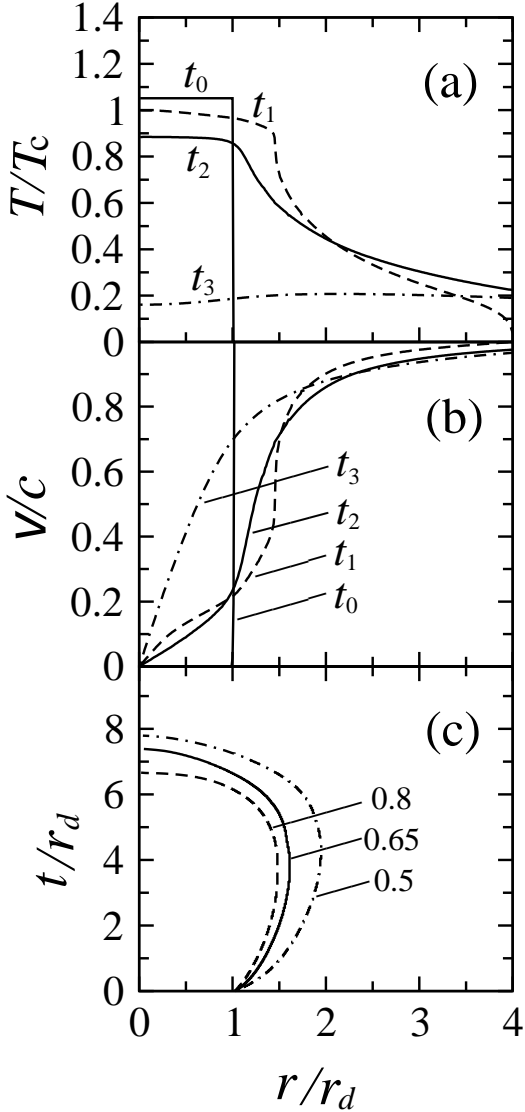}
\caption{\label{fig:f1} (a) Temperature profile and (b) velocity profile for the 
droplet at $t_n = 3n\lambda r_d, \lambda=0.99$. (c) Isotherms for 
the droplet. }
\end{figure}

The two-particle Bose-Einstein correlation function is defined as 
the ratio of the two-particle momentum distribution $P(k_1,k_2)$ to 
the product of the single-particle momentum distribution $P(k_1)
P(k_2)$.  For a chaotic pion-emitting source, $P(k_i)~(i=1,2)$ 
and $P(k_1,k_2)$ can be expressed as \cite{Won94}
\begin{eqnarray}
\label{pk1}
P(k_i) = \sum_{X_i} A^2(k_i,X_i) \,,
\end{eqnarray}
\begin{eqnarray}
\label{pk12}
P(k_1,k_2) = \sum_{X_1, X_2} \Big|\Phi(k_1, k_2; X_1, 
X_2 )\Big|^2 ,
\end{eqnarray}
where $A(k_i,X_i)$ is the magnitude of the amplitude for emitting a
pion with 4-momentum $k_i=(\bsk_i,E_i)$ in the laboratory frame at
$X_i$ and is given by the Bose-Einstein distribution in the local rest
frame of the source point.  $\Phi(k_1, k_2; X_1, X_2 )$ is the
two-pion wave function.  Neglecting the absorption of the emitted
pions by other droplets, $\Phi(k_1, k_2; X_1, X_2 )$ is simply
\begin{eqnarray}
\label{PHI}
& &\!\!\!\!\!\!\!\Phi(k_1, k_2; X_1, X_2 )
\nonumber \\
& & ={ 1 \over \sqrt{2}} \Big[ A(k_1, X_1) 
A(k_2, X_2) e^{i k_1 \cdot X_1 + i k_2 \cdot X_2} 
\nonumber \\
& & ~~~+ A(k_1, X_2) A(k_2, X_1)
e^{i k_1 \cdot X_2 + i k_2 \cdot X_1 } \Big] .~~~~~
\end{eqnarray}
Using the components of ``out" and ``side"
\cite{Pra90,Ber88,Wie99,Wei02} of the relative momentum of the two
pions, $q=|{\bf k_1}-{\bf k_2}|$, as variables, we can construct the
correlation function $C(q_{\rm out},q_{\rm side})$ from $P(k_1,k_2)$
and $P(k_1)P(k_2)$ by summing over ${\bf k_1}$ and ${\bf k}_2$ for
each $(q_{\rm out},q_{\rm side})$ bin.  The HBT radius $R_{\rm out}$
and $R_{\rm side}$ can then be extracted by fitting the calculated
correlation function $C(q_{\rm out},q_{\rm side})$ with the following
parametrized correlation function
\begin{equation}
\label{cq}
C(q_{\rm out},q_{\rm side})=1+\lambda \, e^{-q_{\rm out}^2 R_{\rm out}^2 
-q_{\rm side}^2 R_{\rm side}^2} \,.
\end{equation}

The explicit procedure for calculating the two-pion correlation
function is as follows.

Step 1: select the emission points of the two pions randomly on the 
space-time freeze-out surfaces of the droplets, and get their space-time 
coordinate $X_1$ and $X_2$ in the laboratory frame. 

Step 2: generate the momenta $k'_1$ and $k'_2$ of the two pions in
local frame according the Bose-Einstein distribution characterized by
the temperature $T_f$, and obtain their momenta $k_1$ and $k_2$ in the
laboratory frame by Lorentz transforms.

Step 3: calculate $[E'_1/E_1] [E'_2/E_2]$ for $P(k_1)P(k_2)$ and
$[E'_1/E_1] [E'_2/E_2] \cos[(k_1\!-\!k_2) (X_1\!-\!X_2)]$ for
$P(k_1,k_2)$, and accumulate them in the corresponding $(q_{\rm out},
q_{\rm side})$ bin.

Step 4: repeat steps 1 through 4 many times to get the correlation 
function within a certain accuracy. 

\begin{figure}
\includegraphics{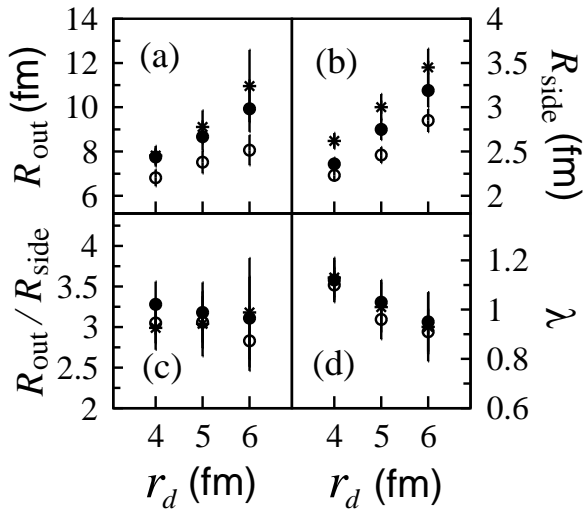}
\caption{\label{fig:f2} The two-pion interferometry results 
for the one-droplet source as a function of the initial droplet radius $r_d$. 
The symbols $\circ$, $\bullet$, and $\star$ are for the freeze-out 
temperatures $T_f = 0.80 T_c$, $T_f = 0.65 T_c$, and $T_f = 0.50 T_c$, 
respectively.}
\end{figure}

We first examine the two-pion correlation function for a singlet
droplet source.  By fitting the two-dimension correlation function
$C(q_{\rm out},q_{\rm side})$ obtained with the above steps with
Eq. (\ref{cq}), we get the parameters $R_{\rm out}$, $R_{\rm side}$,
and $\lambda$ simultaneously.  In our calculations, the transverse
momenta of the pions are integrated over.  The average transverse
momenta of the pions in our fitting samples are 307, 329, and 386 MeV
for $T_f=$0.8$T_c$, 0.65$T_c$, and 0.5$T_c$, respectively.  The reason
for a larger average transverse momentum to associate with a smaller
freeze-out temperature is due to the larger average expansion velocity
in the case of a smaller freeze-out temperature.  Figure 2(a), 2(b),
2(c), and 2(d) show the HBT results $R_{\rm out}$, $R_{\rm side}$,
$R_{\rm out} / R_{\rm side}$, and $\lambda$ as a function of the
initial radius $r_d$ of the droplet, for the freeze-out temperatures
$T_f = 0.80 T_c$ (symbol $\circ$), $T_f = 0.65 T_c$ (symbol
$\bullet$), and $T_f = 0.50 T_c$ (symbol $\star$).  It can be seen
that the HBT radii $R_{\rm out}$ and $R_{\rm side}$ increase linearly
with $r_d$, but the ratio $R_{\rm out} / R_{\rm side}$ is about 3
within the errors.  The radius $R_{\rm side}$ reflects the spatial
size of the source and the radius $R_{\rm out}$ is related to the
lifetime of the source \cite{Pra90,Ber88,Wie99,Wei02}.  From the
hydrodynamical solution in figure 1(c), both the average freeze-out
time and freeze-out radial distance increase with $r_d$ for different
$T_f$.  As a consequence, $R_{\rm out} / R_{\rm side}$ is insensitive
to the values $r_d$ and $T_f$.  The value of $R_{\rm out} / R_{\rm
side} \sim 3$ for a single droplet is however much larger than the
observed values \cite{PHE02,STA01}.  In our calculations, we did not
including resonances in the hadronic phase.  If we take the
hypothetical case of $d_Q/d_H =3$ to include a resonance gas in the
hadronic phase, as discussed by Rischke and Gyulassy \cite{Ris96}, we
find that the ratio of $R_{\rm out} / R_{\rm side}$ is about 2.75.  It
is still much larger than the observed values \cite{PHE02,STA01}.  In
Fig. 1 (d), the values of $\lambda$ for $r_d=4$ fm are larger than
unity.  This is a non-Gaussian correlation function effect, and the
effect is larger for a wider correlation function, corresponding to a
smaller source.

As the average freeze-out time is proportional to the initial radius
of the droplet, the freeze-out time and $R_{\rm out}$ decreases if the
initial radius of the droplet decreases.  On the other hand, $R_{\rm side}$
increases if the width of the droplet spatial distribution increases.
A variation of the droplet size and the width of droplet spatial
distribution can result in $R_{\rm out}$ nearly equal to $R_{\rm
side}$.  Accordingly, we calculate next the two-pion correlation
function for a Gaussian distribution source of $N_d$ droplets.  The
spatial center-of-mass coordinates ${\bf X}_d$ of the droplets are
assumed to obey a static Gaussian distribution$\sim \exp(-{\bf
X}_d^2 / 2 R_{\rm G}^2)$ \cite{Pra92,Zha95,Zha00}.  Figure 3(a), 3(b),
3(c), and 3(d) give the HBT $R_{\rm out}$, $R_{\rm side}$, $R_{\rm out}
/ R_{\rm side}$, and $\lambda$ as a function of the number of droplets
$N_d$ for different values of $r_d$.  In this calculation, we take
$T_f =0.65 T_c$ and $R_{\rm G} = 5.0$ fm.  The symbols $\circ$,
$\bullet$, and $\star$ correspond to $r_d=2.0$ fm, $r_d=1.5$ fm, and
$r_d=1.0$ fm, respectively.  It can be seen that the radii $R_{\rm
out}$ and $R_{\rm side}$ have a slowly increasing tendency as $N_d$
increases but their ratio $R_{\rm out}/ R_{\rm side}$ is almost
independent of $N_d$.  $R_{\rm out}$ decreases as $r_d$ decreases but
$R_{\rm side}$ is relatively independent of $r_d$.  Consequently, the
ratio $R_{\rm out} / R_{\rm side}$ decreases when $r_d$ decreases.
The ratio $R_{\rm out} / R_{\rm side}$ for $r_d=1.5$ fm is about 1.15
which is much smaller than the result of about 3 for the single
droplet source.

\begin{figure}
\includegraphics{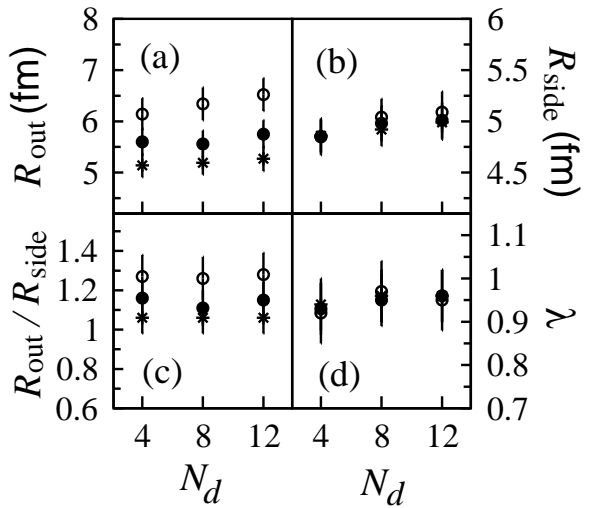}
\caption{\label{fig:f3} Two-pion HBT results for the granular source of 
the droplets.  The symbols $\circ$, $\bullet$, and $\star$ are for the 
initial droplet radius $r_d=$2.0, 1.5, and 1.0 fm, respectively.}
\end{figure}

\begin{figure}
\includegraphics{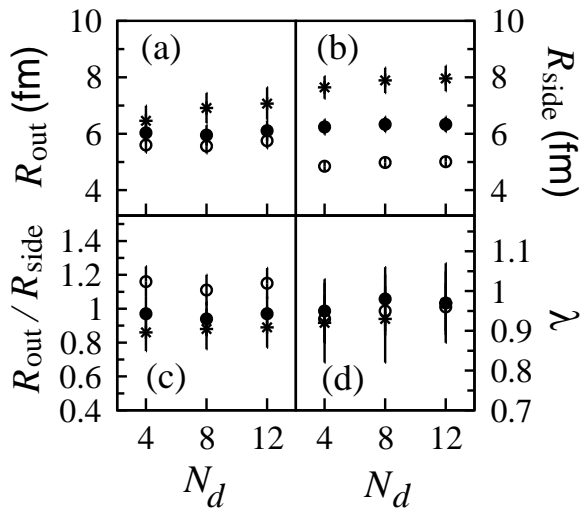}
\caption{\label{fig:f4} Two-pion HBT results for the granular source 
with a radial collective expansion of the droplets. The symbols $\circ$, 
$\bullet$, and $\star$ are for the expansion velocity $v_d=$0, 
0.3, and 0.6, respectively.}
\end{figure}

Finally, to study the effect of additional collective expansion of the
droplets, we calculate the two-pion correlation function for an
expanding source.  The initial distribution of the droplets is the
same Gaussian distribution as the above granular source, but the
droplets are assumed to expand collectively with a 
constant radial velocity
$v_d$ after the initial time, in addition to their hydrodynamical
expansion.  Figure 4(a), 4(b), 4(c), and 4(d) give the HBT $R_{\rm
out}$, $R_{\rm side}$, $R_{\rm out} / R_{\rm side}$, and $\lambda$ for
different values of $v_d$.  In this calculation, we take the initial
radius of the droplets to be $r_d = 1.5$ fm, the other parameters are
the same as the above calculations for a static granular source.  The
symbols $\circ$, $\bullet$, and $\star$ correspond to $v_d=$0, 0.3,
and 0.6, respectively.  The results in Figs. 4(a) and 4(b) show that
$R_{\rm side}$ increases more rapidly with the droplet collective
expansion velocity $v_d$ than $R_{\rm out}$.  A radial expansion will
increase the transverse size of the granular source and $R_{\rm
side}$.  On the other hand, $R_{\rm out}$ measures the source life
time and the spatial extension where the two pions are emitted with
nearly parallel and equal momenta, and the additional radial boost
modifies only slightly the spatial separation of these points for most
cases.  As a result, $R_{\rm out}$ does not increase as rapidly as
$R_{\rm side}$ and $R_{\rm out}/R_{\rm side}$ is smaller at large
$v_d$ than at zero $v_d$ (see Fig 1(c)).  The ratio $R_{\rm
side}/R_{\rm out}$ is of order 1 which is close to the observed value
\cite{PHE02,STA01}.

In summary, we propose a simple model of granular source of
quark-gluon plasma droplets to examine the HBT interferometry data.
The droplets evolve hydrodynamically and pions are emitted thermally
from the droplets at the freeze-out configuration characterized by a
freeze-out temperature $T_f$.  As the average freeze-out time is
proportional to the radius of the droplet, smaller droplet size allows
pions to be emitted within a shorter time and the life-time of the
source decreases, leading to a smaller HBT radius $R_{\rm out}$.  On
the other hand, the HBT radius $R_{\rm side}$ depends on the width of
the spatial distribution of the droplets and is insensitive to the
initial size of the droplets. The ratio of $R_{\rm out}$ to $R_{\rm
side}$ decreases significantly if the emitted source is granular in
nature.  Furthermore, $R_{\rm side}$ increases with the
collective-expansion velocity of the droplets more rapidly than
$R_{\rm out}$.  The ratio $R_{\rm out}/R_{\rm side}$ is about
1.15---0.88 for the collective-expansion velocity of the droplets from
zero to 0.6, for the granular source with a Gaussian initial radius 5
fm and a droplet initial radius 1.5 fm.  This $R_{\rm out}/R_{\rm
side}$ ratio is close to the experimental values \cite{PHE02,STA01}.
The granular model of quark-gluon plasma may provide a possible
explanation to the RHIC HBT puzzle.  It may also lead to a new insight
into the dynamics of the quark-gluon plasma phase transition as the
formation of a granular structure is expected to occur in a
first-order QCD phase transition
\cite{Wit84,Cse92,Ven94,Ala99,Cse03,Ran04}.

In order to bring out the most important features, we have neglected
the multiple scattering effects on HBT interferometry
\cite{Zha04,Won03,Won04}, and have not considered how the granular nature 
of the plasma may arise from detailed phase transition dynamics
\cite{Wit84,Cse03,Ran04}.  The sizes of the droplets in a collision
can also have a distribution.  Future refinements of the present model
to take into account these effects on $R_{\rm out}/R_{\rm side}$ will
be of great interest.

WNZ would like to thank Drs. T. Barnes, V. Cianciolo, and G. Young for
their kind hospitality at Oak Ridge National Laboratory.  This
research was supported by the National Natural Science Foundation of
China under Contract No.10275015 and by the Division of Nuclear
Physics, US DOE, under Contract No. DE-AC05-00OR22725 managed by
UT-Battle, LC.

\end{document}